\documentstyle[11pt,newpasp,twoside,epsf]{article}
\def\edcomment#1{\iffalse\marginpar{\raggedright\sl#1\/}\else\relax\fi}
\reversemarginpar

\begin{document}
\title{The Red-Sequence Cluster Survey:The Value of $\Omega_m$ and
$\sigma_8$} \author{Michael D. Gladders} \author{H.K.C. Yee}
\affil{Department of Astronomy, University of Toronto, 60 St. George
Street, Toronto, ON, Canada, M5S 3H8}

\begin{abstract}
The Red-Sequence Cluster Survey (RCS) is a 100 deg$^2$ optical survey
for high-redshift galaxy clusters. One of the goals of the survey is a
measurement of $\Omega_m$ and $\sigma_8$ via the evolution of the mass
spectrum of galaxy clusters. Herein we briefly describe how this will
initially be done, and also demonstrate the eventual power of the RCS
for this type of measurement by a qualitative analysis of the first
1/10th of the survey data.
\end{abstract}

\section{The RCS and Cosmology}
\paragraph{}
The RCS will provide a large sample of optically selected galaxy
clusters over the redshift range $0<z<1.4$. Clusters are selected
using a technique based on locating the red-sequence of early-type
galaxies in the cluster core (Gladders \& Yee 2000a). Simulations show
that the RCS will be complete for clusters of line-of-sight velocity
dispersions of 750 km~sec$^{-1}$ to $z\sim1.1$ (Gladders \& Yee
2000b), regardless of the details of the cluster galaxy properties.
\paragraph{}
The large and well-understood cluster sample from the RCS will be well
suited to measuring the evolution of the cluster mass spectrum,
$N(M,z)$, and hence $\Omega_m$ and $\sigma_8$.  Individual cluster
redshifts can be estimated photometrically from the survey data to an
accuracy of $\sim7\%$ (Gladders \& Yee 2000a). Masses can also be
estimated photometrically, using the $B_{gc}$ richness estimator,
which has been shown to be well correlated with velocity dispersion
(Yee \& L\'{o}pez-Cruz 1999) and hence mass. The $B_{gc}$ - mass
correlation is currently calibrated to $z\sim0.6$; once calibration is
complete over the entire RCS redshift range the whole survey should
yield estimates of $\Omega_m$ and $\sigma_8$ accurate to $\sim5\%$ -
more accurate and independent of current results based on relatively
small X-ray selected samples (e.g., see Henry 2000 and references
therein).

\paragraph{}
As a qualitative illustration of the cosmology indicated by the RCS,
we have used the RCS selection functions (Gladders \& Yee 2000a,b) to
integrate out the mass dependence in model predictions of $N(M,z)$ and
compared the RCS measure of $N(z)$ to predictions from two typical
cosmologies. Predictions of $N(M,z)$ are made from the standard
Press-Schechter formalism, and multiplied by the RCS selection
functions (expressed in mass and redshift, assuming a cluster with a
typical luminosity function shape, galaxy mixture and cluster shape,
concentration and size). These selection functions tail off to zero
probability at a lower mass limit which becomes progressively more
massive at higher redshift, limiting the contribution to
$N(M,z)$ from lower-mass clusters and groups.  The result of these
computations, as well as the actual counts from a portion of the RCS,
are shown in Figure 1. Clearly, and not surprisingly, the low
$\Omega_m$ and high $\sigma_8$ model is preferred. Regardless of the
precise details of the shapes of the modeled curves (which are
dependent on knowing the selection functions), the simple abundance of
candidates at high redshift in the real data is a strong argument for
a low $\Omega_m$ universe.

\begin{figure}[!h] 
\plotfiddle{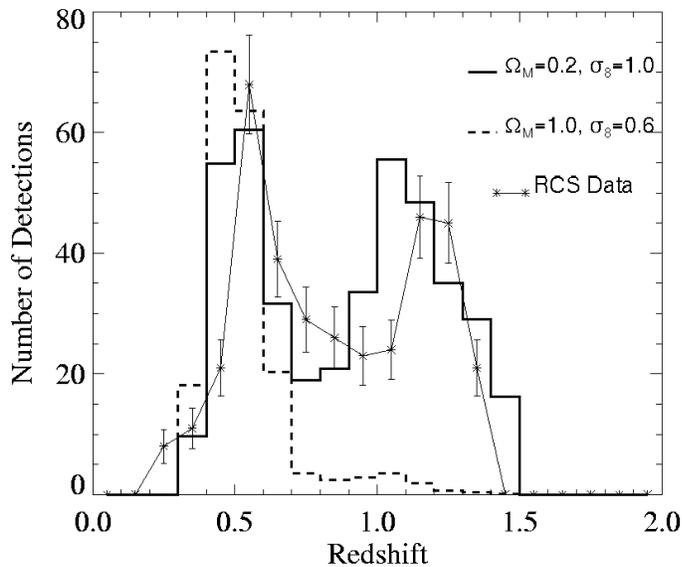}{7cm}{0}{50}{50}{-175}{-100}
\caption{The predicted counts for clusters and groups are shown for
two different cosmologies (thick lines) as well as the measured counts
of significant detections in a subset of the RCS (thin line +
asterisk).  The low $\Omega_m$ model is vastly preferred. The
apparent excess of low redshift objects (resulting in an apparent
`double-peaked' appearance) is the result of redshift aliasing from
lower redshifts, and the great sensitivity of the RCS to even $300$
km~s$^{-1}$ groups at moderate redshifts (Gladders \& Yee 2000a).  }
\end{figure}


\begin{references}
\reference Gladders, M.D., \& Yee, H.K.C. 2000a, AJ, 120, 2148
\reference Gladders, M.D., \& Yee, H.K.C. 2000b, to be submitted to AJ
\reference Henry, J.P. 2000, ApJ, 534, 565
\reference Yee, H.K.C, \& L\'{o}pez-Cruz, O. 1999, AJ, 117, 1985

\end{references}
\end{document}